\documentclass[prx,amsmath,amssymb,aps,floatfix,longbibliography,nofootinbib]{revtex4-2}

\usepackage{graphicx}
\usepackage{dcolumn}
\usepackage{bm}
\usepackage[T1]{fontenc}
\usepackage[dvipsnames]{xcolor}
\newcommand{\authormain}{Farshid Jafarpour}
\newcommand{\titlemain}{Exactly Solvable Population Model with Square-Root Growth Noise and Cell-Size Regulation}
\usepackage[bookmarks={false}, pdfauthor={\authormain}, pdftitle={\titlemain}]{hyperref}
\hypersetup{colorlinks=true, linkcolor=BrickRed, citecolor=Violet, filecolor=OliveGreen, urlcolor=RoyalBlue, filebordercolor={.8 .8 1}, urlbordercolor={.8 .8 0}}

\usepackage{orcidlink}
\newcommand{\Ei}{\operatorname{Ei}}

\newcommand{\mean}[1]{\left\langle#1\right\rangle}
\newcommand{\cmean}[2]{\left\langle\left.#1\right|#2\right\rangle}
\newcommand{\Var}[1]{\text{Var}\left({#1}\right)}
\newcommand{\prob}[1]{\text{{Prob}}\left({#1}\right)}

\newcommand{\eq}[1]{Eq.~\eqref{eq:#1}}
\newcommand{\Eq}[1]{Equation~\eqref{eq:#1}}
\newcommand{\fig}[1]{Fig.~\ref{fig:#1}}
\newcommand{\Fig}[1]{Figure~\ref{fig:#1}}

\newcommand{\apx}[1]{Appendix~\ref{sec:#1}}
\newcommand{\Sec}[1]{Section~\ref{sec:#1}}

\begin{document}

\title{\titlemain}

\author{Farshid Jafarpour\,\orcidlink{0000-0002-2441-0296}}
\affiliation{Institute for Theoretical Physics, Department of Physics, Utrecht University, Utrecht, Netherlands}
\affiliation{Centre for Complex Systems Studies, Utrecht University, Utrecht, Netherlands}
\email{f.jafarpour@uu.nl}

\date{\today}

\begin{abstract}
Stochastic exponential growth is nearly ubiquitous across cellular life, but how its microscopic noise structure shapes population growth remains poorly understood. Here, we introduce an exactly solvable population model in which cells grow exponentially with fluctuations that scale with the square root of cell size, and divide according to general size-control mechanisms. Our first result is that the population growth rate is exactly equal to the mean single-cell growth rate, for all noise strengths and for all division and size-regulation schemes that maintain size homeostasis. Thus square-root growth noise does not affect long-term fitness, in sharp contrast to models with size-independent stochastic growth rates. 
Second, we derive an exact solution for the steady-state distribution of cell sizes in the population and show that it is broadened by growth fluctuations.
Third, the mean-rescaled population size $N_t/\mean{N_t}$ converges to a stationary compound Poisson-exponential distribution that depends only on growth noise. This distribution, and hence the long-time shape of population-size fluctuations, is unchanged by division-size noise or asymmetric partitioning. These results identify Feller-type exponential growth with square-root noise as an exactly solvable benchmark for stochastic growth in size-controlled populations and provide concrete signatures that distinguish it from models with size-independent growth-rate noise.
\end{abstract}

\maketitle

\section{Introduction}

How microscopic growth fluctuations shape population fitness and size statistics remains a central unresolved problem in quantitative cell physiology. Over the past decade, a broad suite of single-cell technologies has enabled long-term measurements of growth and division for individual cells under tightly controlled conditions~\cite{wang2010robust, falconnet2011microfluidic, son2012direct, allard2022microfluidics}. These data have enabled a detailed characterization of division statistics and cell-size control strategies across species and growth conditions~\cite{campos2014constant, amir2014cell, taheri2015cell, Soifer2016CommonStrategy, si2019mechanistic, luo2021master, elgamel2024effects, elgamel2025general, genthon2025noisy, genthon2026cell}. By contrast, the temporal statistics of single-cell growth rate remain poorly constrained. To discriminate between them, one needs analytically tractable population models that connect microscopic growth fluctuations to population-level observables in a way that allows different growth mechanisms to be distinguished.

Early characterization of growth variability was limited to the observation that the average rate of growth throughout the cell cycle varies from cell to cell and relaxes to an average value over a few generations~\cite{taheri2015cell, iyer2014scaling, cadart2018size}, while some attempts have been made to characterize instantaneous growth fluctuations with limited resolution~\cite{kiviet2014stochasticity, cadart2022volume, levien2026stochasticity}. Many current theoretical treatments assume that cell size $s(t)$ grows exponentially,
\begin{equation}
    \frac{ds}{dt} = \lambda(t)\,s,
\end{equation}
with an instantaneous exponential growth rate $\lambda(t)$ that fluctuates in time but is otherwise independent of the cell size $s$. Within this broad class, explicit solutions have been obtained for various growth-rate processes $\lambda(t)$, including models with independent generation times and models in which $\lambda(t)$ is an Ornstein-Uhlenbeck or related stationary process~\cite{tuanase2008regulatory, thomas2018sources, levien2021non, hein2024competition, hein2024asymptotic, levien2025size}. These analyses show that growth-rate fluctuations can substantially modify the population growth rate $\Lambda$ relative to single-cell averages but do not perturb snapshot cell-size distributions. In particular, for size-independent $\lambda(t)$, the population growth rate generally depends on the full statistics of $\lambda(t)$ and increases with its variance and correlation time under broad conditions~\cite{tuanase2008regulatory, levien2021non, hein2024asymptotic}. Recent decoupling theorems further show that, for this class of models, $\Lambda$ is determined solely by the growth process, whereas the steady-state cell-size distribution is determined solely by the statistics of division and size regulation~\cite{hein2024asymptotic, levien2025size}. However, all of these results rely on the assumption that the noise amplitude in the growth dynamics does not depend explicitly on cell size.

From a microscopic perspective, that assumption is not a natural expectation for intrinsic fluctuations: larger cells contain more molecules and should exhibit smaller relative fluctuations\footnote{Since the deterministic exponential-growth law already treats biomass production as an extensive process, the corresponding intrinsic production noise is expected to have an extensive variance, and hence a noise amplitude proportional to the square root of cell size.}. Coarse-graining discrete birth-death or autocatalytic reaction networks for cell mass leads, in the diffusion limit, to Langevin equations in which the noise amplitude scales as the square root of size, rather than as a size-independent prefactor~\cite{VANKAMPEN2007244, iyer2014universality, pirjol2017phenomenology}. This limiting dynamics is the Feller square-root diffusion, which preserves non-negativity of the size variable and stationary mean-rescaled distributions~\cite{feller1951two,pirjol2017phenomenology}. These theoretical observations, in combination with experimental single-cell analyses that suggest fluctuations in the instantaneous exponential growth rate $d\ln s/dt$ decay as cells progress through the cell cycle~\cite{cadart2022volume, biswas2024collective} motivate the study of population dynamics under stochastic exponential growth with square-root noise.

\begin{figure}
    \centering
    \includegraphics[width=\linewidth]{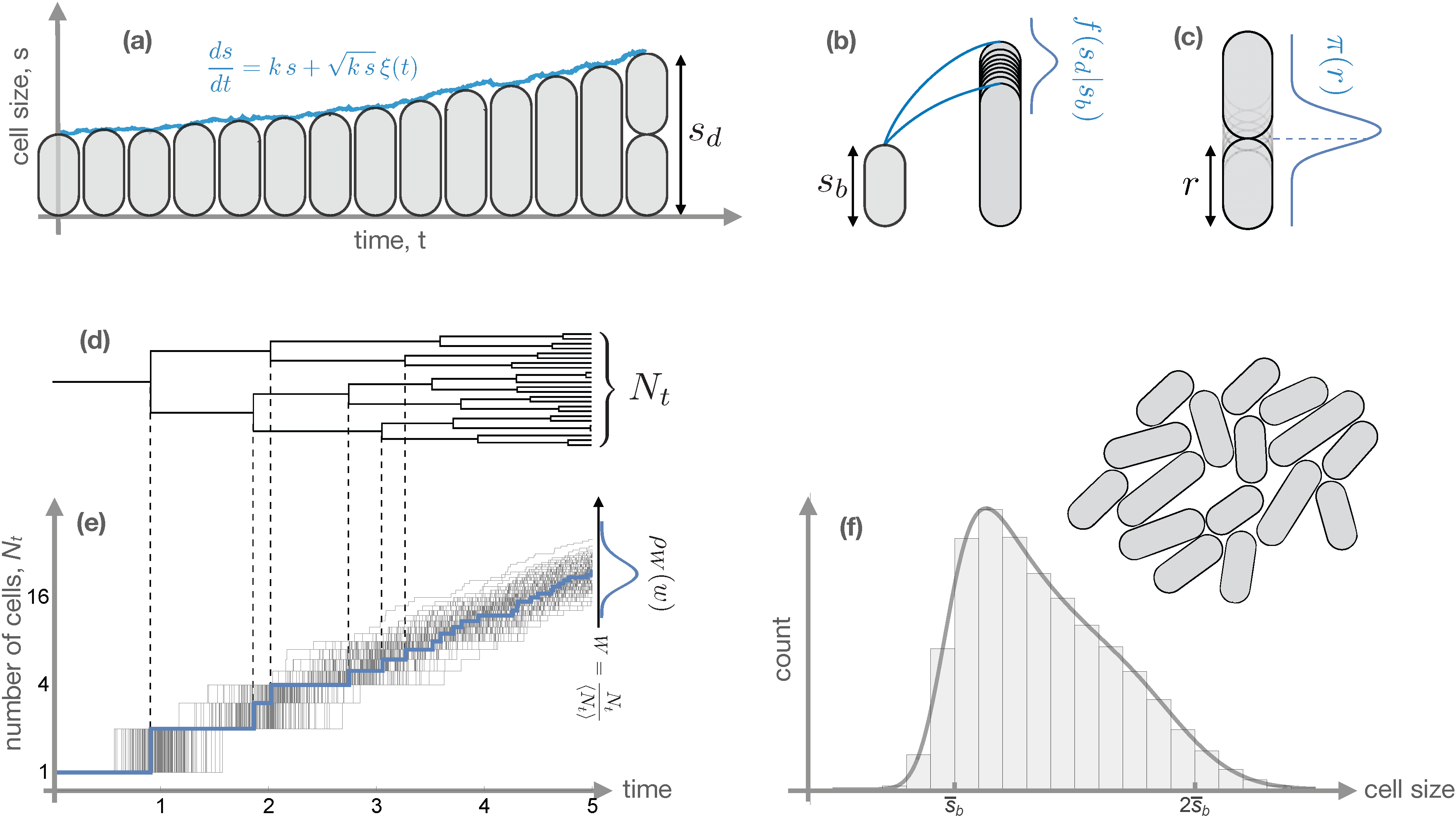}
    \caption{Schematic representation of the single-cell model of growth and division (top row) and the resulting population properties (bottom half). (a) Sample trajectory of the size $s$ of an individual cell grows according to a stochastic exponential growth model with square root noise. (b) The division size $s_d$ is a random variable, with its probability density $f(s_d|s_b)$ depending on the birth size $s_b$. Particular choices of this conditional density and its $s_b$ dependence reduce to known models of cell-size regulation. (c) At division, two daughter cells are born with sizes $s_b = r\, s_d$ and $s_b' = (1-r) s_d$ with the ratio $r$ drawn from a distribution $\pi(r)$ with mean $0.5$. (d) A sample population tree showing all the division events until a given time. (e) Log-scale plot of the total number $N_t$ of cells as a function of time for several populations with the same initial conditions (thin gray lines). The trajectory corresponding to the population tree above is highlighted (thick blue line) with jumps corresponding to division events. The asymptotic slope of these curves is the population growth rate. The mean-rescaled distribution $\rho_W(w)$ of the cell counts is shown for the final time. (f) The population cell-size distribution can be obtained by histogramming the sizes of all the cells in the population at a given time. Analytical solutions for the population growth rate, mean-rescaled population number distribution, and population cell-size distribution are the main results of this paper.}
    \label{fig:schematics}
\end{figure}

In this work, we analyze a size-structured population in which individual cells grow according to a Feller square-root process,
\begin{equation}
    \frac{ds}{dt} = k\,s + \sigma\,\sqrt{2k\,\bar s_b\, s}\,\xi(t),
    \label{eq:intro_feller}
\end{equation}
where $k$ is the mean exponential growth rate, $\bar s_b$ is the mean birth size, $\sigma$ is a dimensionless measure of growth fluctuations, and $\xi(t)$ is standard Gaussian white noise interpreted in the It\^o sense. Here, we ignore any temporal correlations in the growth rate (colored-noise extensions are discussed in \Sec{discussion}). Division and size regulation are modeled at the level of division size and partitioning statistics and are otherwise left arbitrary, provided they yield a stationary lineage distribution of birth sizes (see \fig{schematics} (a-c)). Despite the apparent complexity of this model in combination with the generality of cell-size regulation, we obtain a series of exact results for the population growth rate, snapshot cell-size distributions, and population-size fluctuations (see \fig{schematics} (d-f)).

Our first main result is that the asymptotic population growth rate
\begin{equation}
    \Lambda = \lim_{t\to\infty} \frac{d}{dt}\ln N_t
\end{equation}
is exactly equal to the deterministic single-cell growth rate, $\Lambda = k$, for all $\sigma$ and for all models of division and size-regulation compatible with size homeostasis. This is in sharp contrast to models with size-independent $\lambda(t)$, where $\Lambda$ depends sensitively on the statistics of $\lambda(t)$~\cite{tuanase2008regulatory, levien2021non, hein2024asymptotic}. It is also notable because the Feller process~\eqref{eq:intro_feller} does not satisfy the structural assumptions underlying recent general decoupling results~\cite{levien2025size}, yet it exhibits an even stronger form of decoupling: the population growth rate is completely insensitive to both growth noise and division. As shown below, this stronger decoupling follows from the additive square-root noise and should not be interpreted as a generic property of stochastic growth models.

Our second set of results concerns the steady-state snapshot distribution of cell sizes in the population. We show that growth noise modifies this distribution in a simple way: once the deterministic growth $\sigma=0$ solution is known, the full $\sigma>0$ solution is obtained by a one-sided exponential convolution with a kernel of width $\sigma^2$. Thus, division noise, asymmetric division, and size regulation enter only through the deterministic solution, while growth noise introduces a division-independent smoothing. In particular, the population mean size is shifted by $-\sigma^2$, whereas the variance is modified only at order $\sigma^4$ relative to the deterministic growth solution. The population cell-size distribution for the deterministic growth model, $\sigma=0$ has a known closed form in terms of the lineage birth-size distribution.

Finally, we analyze fluctuations in the total number of cells $N_t$. We compute the full asymptotic distribution of the mean-rescaled population size,
\begin{equation}
    W \equiv \lim_{t\to\infty} \frac{N_t}{\mean{N_t}},
\end{equation}
and show that this distribution has a closed form that depends only on $\sigma$. In particular, $\Var{N_t}/\mean{N_t}^2 = 2\sigma^2$ at long times, independent of division and size regulation. To the best of our knowledge, this is the only nontrivial branching model, beyond the Yule process, for which the full stationary distribution of $W$ is known in closed form~\cite{bellman1952age}. Importantly, division-size noise, partitioning noise, and size regulation only modify the prefactor relating total mass to total cell number; the mean-rescaled distribution $W$ itself remains unchanged. The shape of population-size fluctuations is therefore completely independent of the division mechanism and the details of size control.

Taken together, these results identify stochastic exponential growth with square-root noise as an exactly solvable benchmark for growth fluctuations in size-structured populations. The model reveals a hierarchy of decouplings: (i) the population growth rate $\Lambda$ is independent of growth noise and division; (ii) the snapshot cell-size distribution depends on cell-size regulation only through the stationary lineage birth-size distribution and is modified by growth noise through an exponential convolution; and (iii) the distribution of the mean-rescaled population size is fully determined by the growth noise parameter $\sigma$, independent of the division mechanism and size regulation. Because these predictions differ qualitatively from those of models based on size-independent growth rate fluctuations, they provide concrete, testable signatures that can be used to distinguish competing microscopic models of growth using existing and future single-cell data.

The remainder of this paper is structured as follows. We first show in \Sec{pop_growth_rate} that the population growth rate is equal to the mean single-cell growth rate. In \Sec{det_div_size}, we study the special case where cells divide symmetrically at a fixed division size. For this case, we derive the steady-state population cell-size distribution in \Sec{det_size_dist}. We then show in \Sec{Bellman_Harris} that the corresponding population dynamics can be viewed as a Bellman--Harris branching process and derive the division time distribution as first-passage times of the Feller process. \Sec{number_fluctuation} derives the distribution of the total number of cells. In \Sec{div_noise}, we allow noisy division sizes and general size-regulation mechanisms, showing that they enter the population cell-size distribution only through the deterministic-growth solution. In \Sec{asym_div}, we include stochastic partitioning and asymmetric division, showing that the same convolution structure persists. We conclude by discussing the implications and possible extensions of these results in \Sec{discussion}.

\section{Population Growth Rate}\label{sec:pop_growth_rate}

A defining feature of the square root process in \eq{intro_feller} is its additivity, which greatly simplifies its population behavior. In particular, the total population mass $M_t\equiv\sum_i s_i$ obeys the same equation as the single cell size
\begin{equation}\label{eq:s_tot}
    \frac{dM_t}{dt} = k\, \sum_i s_i + \sigma\sum_i \sqrt{2\,k\,\bar s_b\, s_i}\, \xi_i(t) = k\, M_t+\sigma \sqrt{2\,k\,\bar s_b\, M_t}\, \xi(t).
\end{equation}
Here we have used the fact that the single-cell noises $\{\xi_i(t)\}$ are independent standard Gaussian white noises. Thus the sum $\sum_i \sqrt{s_i}\,\xi_i(t)$ is again a Gaussian white noise with variance $\sum_i s_i = M_t$, which we rewrite as $\sqrt{M_t}\,\xi(t)$ for a new standard white noise $\xi(t)$. Division is assumed to be mass-conserving, so the total mass $M_t$ is unaffected by division events.

Asymptotically, the noise term grows more slowly than the deterministic term, giving rise to deterministic exponential growth. At the long time limit, when cell sizes are regulated, the total number $N_t$ of cells is given by the total population mass $M_t$ divided by the population average of cell sizes $\bar s_p$. Using It\^o's lemma, the asymptotic population growth rate is given by
\begin{equation}
    \Lambda = \lim_{t\to\infty} \frac{d}{dt}\ln(N_t) = \lim_{t\to\infty} \frac{d}{dt}\ln\left(\frac{M_t}{\bar s_p}\right) =  \lim_{t\to\infty}\left(k - \frac{\sigma^2\, k\, \bar s_b}{M_t} +\sigma\,\sqrt{\frac{2\,k\,\bar s_b}{M_t}}\,\xi(t) \right)= k
\end{equation}
Unlike other models of growth, where growth fluctuations change the population growth rate~\cite{jafarpour2018bridging, lin2017effects, jafarpour2019cell, lin2020single, tuanase2008regulatory, thomas2018sources, hashimoto2016noise, levien2021non, hein2024asymptotic}, in this model, the population growth rate is exactly the mean single-cell growth rate $k$, independent of the fluctuation strength $\sigma$ or any details of cell division.

Note that biologically, this closure seen in \eq{s_tot} is expected only when the dominant intrinsic growth fluctuations arise from many approximately independent production events. It is not expected when fluctuations are dominated by shared cell-wide variables, environmental fluctuations, or long-lived inherited states that correlate the growth noise of many components of the same cell.

\section{Constant Division Size Model}\label{sec:det_div_size}
We start with the simple case, where there is no noise in the division size. Each cell starts with the size $s=\bar s_b$ and divides when it reaches the division size $s_d=2\bar s_b$. Here and for the rest of the paper, without loss of generality, we set $k=1$ and $\bar s_b=1$ by measuring time and size in units of $1/k$ and $\bar s_b$, respectively. Dimensional results can be recovered by replacing $t\to k t$ and $s\to s/\bar s_b$ in the formulas that follow.

\subsection{Asymptotic Size Distribution} \label{sec:det_size_dist}
We examine the distribution of cell sizes within the population. Let $n(s,t)$ be expected density of cells with size $s$ at time $t$, then $n$ would obey the partial differential equation
\begin{equation}\label{eq:det_size_dist_dyn}
    \partial_t n(s,t) = -\partial_s(s\,n(s,t)) + \sigma^2\partial_{ss}(s \, n(s,t)) + 2\,J(s_d)\,\delta(s-s_b) - \Lambda\,n(s,t),
\end{equation}
for $0<s<s_d=2$ with absorbing boundaries at $s=s_d=2$ (due to division). The first two drift and diffusion terms express the change in density due to stochastic cell growth. In the third term, $J(s,t)\equiv s\,n(s,t) - \sigma^2\partial_s(s\,n(s,t))$ is the net current passing through size $s$. Its value at $s_d$ is the division rate and affects the change in density at the birth size $s=s_b$. The factor 2 comes from binary division. The final term represents the dilution of the density due to population growth at the rate $\Lambda=1$. At the boundary $s=0$, both the drift and the diffusion coefficient vanish, so any trajectory that reaches zero remains there~\cite{feller1951two}. This is a naturally absorbing boundary; no externally imposed boundary condition is needed.

\begin{figure}
    \centering
    \includegraphics[width=0.5\linewidth]{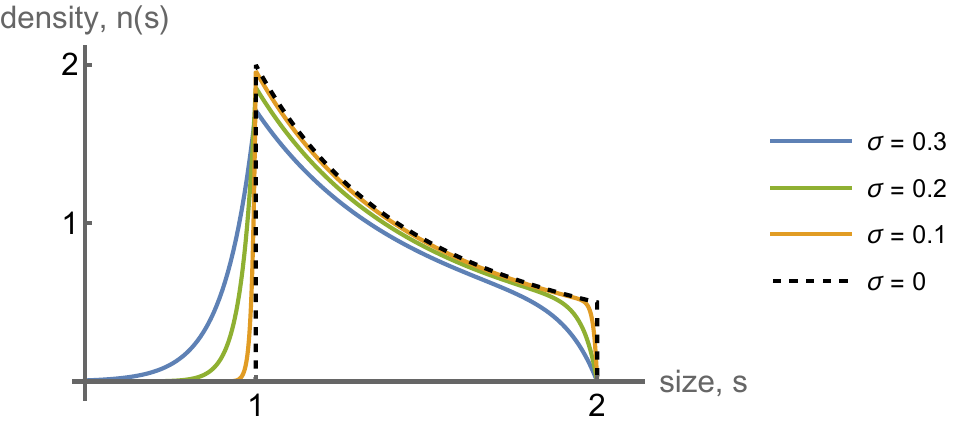}
    \caption{Analytically calculated steady-state cell-size distribution from \eq{det_size_dist} for $\sigma\in\{0.1,0.2,0.3\}$ (solid lines) compared to the cell-size distribution for deterministic growth (dashed line).}
    \label{fig:det_size_dist}
\end{figure}

Let us define $v(s) \equiv s\,n(s)$ at the steady state. For $s\neq1$, $v(s)$ satisfies
\begin{equation}
    \sigma^2s\,v''(s)-s\,v'(s)-v(s)=0,
\end{equation}
with the two independent solutions
\begin{equation}\label{eq:hom_sol}
    v_1(s) = s\,e^{\frac{s}{\sigma^2}},\quad\text{and}\quad v_2(s) = \sigma^2 + s\,e^{\frac{s}{\sigma^2}}\Ei\left(-\frac{s}{\sigma^2}\right),
\end{equation}
where $\Ei(x)\equiv-\int_{-x}^\infty \frac{e^{-t}}{t}dt$ is the exponential integral function. The solutions are linear combinations of $v_1$ and $v_2$ in the intervals $0<s<s_b$ and $s_b<s<s_d$, with four conditions fixing the four coefficients: $v(0)=0$, continuity, $v(s_b^-)=v(s_b^+)$, current matching $2J(s_d) = J(s_b^+)-J(s_b^-)$, and the normalization $n_{\rm abs}+\int_0^{s_d} n(s)ds=1$, where $n_{\rm abs}\equiv -J(0)$ is the mass absorbed at the $s=0$ boundary ($n_{\rm abs}$ is astronomically small for $\sigma \ll 1$). The absorbing boundary $v(s_d)=0$ is automatically satisfied (for $s_d=2s_b$) after imposing these conditions. The final solution for the steady-state size distribution $n(s) = v(s)/s$ is given by
\begin{equation}\label{eq:det_size_dist}
    n(s) = \underbrace{A\,\sigma^2}_{n_{\rm abs}}\,\delta(s)+\begin{cases}
        A\, e^{\frac{s}{\sigma^2}}, & 0 < s \leq s_b\\
        B\, e^{\frac{s}{\sigma^2}}+C\left(\frac{\sigma^2}{s} + e^{\frac{s}{\sigma^2}}\Ei\left(-\frac{s}{\sigma^2}\right)\right), & s_b \leq s < s_d
    \end{cases}
\end{equation}
with
\begin{equation}
\begin{split}
    A &= \frac{2e^{\frac{-s_b}{\sigma^2}}}{\sigma^2} -\frac{e^{\frac{-s_d}{\sigma^2}}}{\sigma^2} +\frac{2s_b}{\sigma^4}\left(\Ei\left(-\frac{s_b}{\sigma^2}\right)-\Ei\left(-\frac{s_d}{\sigma^2}\right)\right),\\
    B &= -\frac{e^{\frac{-s_d}{\sigma^2}}}{\sigma^2} -\frac{2 s_b}{\sigma^4}\Ei\left(-\frac{s_d}{\sigma^2}\right),\qquad \text{and}\qquad C = \frac{2 s_b}{\sigma^4}.
\end{split}
\end{equation}
\Fig{det_size_dist} shows the behavior of \eq{det_size_dist} for $\sigma\in\{0.1,0.2,0.3\}$ as well as the $\sigma\to 0$ limit. For $\sigma=0$, this equation simplifies to the well-known inverse square law $n(s) = 2s_b/s^2$ for $s_b<s<2s_b$ and zero otherwise~\cite{maclean1961some,koch1962model,koch1966distribution}. For bacterial cells in many growth conditions, the growth fluctuations are of the order of $\sigma\sim 0.1$~\cite{taheri2015cell}. As shown in \fig{det_size_dist}, the effect of such low growth fluctuations on size distribution is small. Moreover, the absorption probability $n_{\rm abs}$ at $s=0$ is exponentially small in $1/\sigma^2$ and is completely negligible for biologically relevant values of $\sigma$. For $\sigma=0.3$, the absorption probability at $s=0$ is about $n_{\rm abs} \sim 10^{-6}$, while for $\sigma=0.1$, it is about $n_{\rm abs} \sim 10^{-45}$. 

It is useful to evaluate the mean and variance of cell sizes in the population
\begin{equation}\label{eq:pop_<s>_det}
    \mean{s}_{\rm pop} = 2\ln 2-\sigma^2 + O\left(e^{-1/\sigma^2}\right) \quad\text{and}\quad \Var{s}_{\rm pop} = 2-4(\ln2)^2+\sigma^4 + O\left(e^{-1/\sigma^2}\right).
\end{equation}
The population mean of cell sizes shifts to the left by $\sigma^2$ as we increase $\sigma$. The variance, however, is only affected at order $\sigma^4$.

\subsection{Bellman-Harris Mapping and Generation Time Distribution}\label{sec:Bellman_Harris}
In the absence of the division size variability, the subsequent generation times of the cell are independent of each other, and this model reduces to the case of the independent generation time model (a special case of the Bellman-Harris model~\cite{bellman1952age}), with its division time distribution given by the first passage time of the square root process. All population properties in the Bellman-Harris process are determined by the generation time distribution and its moment generating function, which we find below. These results are used in \Sec{number_fluctuation} to compute the distribution of the number of cells, and in \Sec{sym_birth_rate} where the population division rate depends on the moment generating function of the generation times.

For a cell of size $0 \leq s\leq s_d=2$ at time zero, let $T$ be the time the cell size reaches the division size $s = s_d$.  The survival probability $S(\tau|s) \equiv \prob{T > \tau}$ is the probability that a cell of size $s$ has not divided by time $\tau$ and obeys the backward Fokker-Planck equation
\begin{equation}\label{eq:survival_PDE}
    \partial_\tau S(\tau|s) = s\, \partial_s S(\tau|s) + \sigma^2\, s\, \partial_{ss}S(\tau|s),
\end{equation}
with the boundary conditions
\begin{equation}
    S(\tau|s=0)=1, \quad S(\tau|s=s_d)=0, \text{ and for } s<s_d,\quad  S(0|s) = 1.
\end{equation}
Taking the Laplace transform of \eq{survival_PDE} with respect to $\tau$, we have 
\begin{equation}
    p\,\tilde S(p|s)- \underbrace{S(\tau=0|s)}_{1} = s\, \partial_s \tilde S(p|s) + \sigma^2\, s\, \partial_{ss}\tilde S(p|s),
\end{equation}
where $\tilde S(p|s) \equiv\int_0^\infty S(\tau|s) e^{-p\,\tau}d\tau$. For a given $p$, this is an ODE with the boundary conditions $\tilde S(p|0)=1/p$ and $\tilde S(p|s_d)=0$ with the solution
\begin{equation}
    \tilde S(p|s) = \frac{1}{p}\left(1-\frac{s}{s_d}\frac{{}_1F_1(1-p; 2; -s/\sigma^2)}{{}_1F_1(1-p; 2; -s_d/\sigma^2)} \right)
\end{equation}
where ${}_1F_1$ is the confluent hypergeometric function (also known as Kummer’s function). The probability density $\rho(\tau|s)$ of the waiting time $T$ to division is given by $\rho(\tau|s)=-dS(\tau|s)/d\tau$, and therefore, its Laplace transform is given by $\tilde \rho(p|s) = -p\,\tilde S(p|s)+1$. 

The waiting time for division starting from the birth size $s_b=1$ is the generation time of the cell, and the Laplace transform of its density is given by
\begin{equation}\label{eq:gen_time_laplace}
    \tilde \rho(p) = \frac{s_b}{s_d}\frac{{}_1F_1(1-p; 2; -s_b/\sigma^2)}{{}_1F_1(1-p; 2; -s_d/\sigma^2)},
\end{equation}
with $s_b=1$ and $s_d=2$. The moment generating function $M_T(p)\equiv \mean{e^{p\,T}}$ of the generation time distribution is given by $M_T(p) = \tilde \rho(-p)$, therefore $\tilde \rho(0)$ should be one for a normalized $\rho$. However, since there is a small probability that the cell size hits zero and never divides, our expression for $\tilde \rho$ at zero gives the probability that a cell divides (one minus the probability of absorption at zero)
\begin{equation}
    p_{\rm div} = \tilde \rho(0) = \frac{1-e^{-\sigma^{-2}}}{1-e^{-2\sigma^{-2}}}, \quad\text{and}\quad p_{\rm abs}=1-p_{\rm div} = \frac{e^{-\sigma^{-2}}-e^{-2\sigma^{-2}}}{1-e^{-2\sigma^{-2}}}.
\end{equation}
The lineage absorption probability $p_{\rm abs}$ differs from the steady-state population mass $n_{\rm abs}$ at $s=0$ in \eq{det_size_dist}; both quantities are nevertheless exponentially small in $1/\sigma^2$, for $\sigma\ll 1$. For example, the probability of absorption for $\sigma \sim 0.1$ is about $p_{\rm abs}\sim 10^{-44}$.

For cells that do divide, the moments of their generation times are given by the derivatives of $\tilde\rho(-p)/p_{\rm div}$ at $p=0$. For small $\sigma$, the mean and the variance are given by
\begin{equation}\label{eq:gen_time_mean_var}
    \mean{T} = \ln2 + \frac12\sigma^2 + \frac34 \sigma^4+O(\sigma^6),\quad\text{and}\quad \Var{T} = \sigma^2 + 3\,\sigma^4 + O(\sigma^6).
\end{equation}

In Bellman-Harris, many asymptotic properties of the population are linked to the value of $\tilde \rho(p)$ at integer values of $p$. To obtain these, we can use the following identity
\begin{equation}
    {}_1F_1(1-p; 2; -x) = \sum_{n=0}^{p-1} \binom{p-1}{n}\frac{x^n}{(n+1)!}
\end{equation}
This gives the values of $\tilde \rho(p)$ at integer values of $p=m$ as
\begin{equation}\label{eq:rho_int}
    \tilde \rho(1) = \frac12,\quad \tilde \rho(2) = \frac{1+2\sigma^2}{4+4\sigma^2},\quad \tilde \rho(3)=\frac{1+6\sigma^2+6\sigma^4}{8+24\sigma^2+12\sigma^4},\quad \dots,\quad\tilde \rho(m) = \frac{\sum_{k=0}^{m-1} \binom{m-1}{k}\frac{1}{(k+1)!\,\sigma^{2k}}}{\sum_{k=0}^{m-1} \binom{m-1}{k}\frac{2^{k+1}}{(k+1)!\,\sigma^{2k}}}.
\end{equation}
In particular, the relation $\tilde\rho(1)=1/2$ is exactly the Euler-Lotka equation $\tilde\rho(\Lambda)=1/2$ and confirms the population growth rate $\Lambda=1$. The rest can be used to calculate the moments of the population size, as we will see below. 

\subsection{Distribution of Total Numbers of Cells}\label{sec:number_fluctuation}
For the Bellman-Harris process, the total number of cells $N_t$ is a random variable whose average grows exponentially as $\mean{N_t}= c\,e^{\Lambda t}$ at long time, where the prefactor $c$ can be computed using the Laplace transform of the generation time distribution at $p=\Lambda=1$~\cite{bellman1952age},
\begin{equation}\label{eq:exp_prefactor}
    c=-\frac{1}{4\,\tilde \rho'(1)} =  \left[2\left(\ln2 + \Ei\left(-\frac{1}{\sigma^2}\right) - \Ei\left(-\frac{2}{\sigma^2}\right)\right) - \sigma^2\left(1-e^{-1/\sigma^2}\right)^2\right]^{-1} \approx \frac{1}{2\ln2-\sigma^2} + O\left(e^{-1/\sigma^2}\right)
\end{equation}

The mean-rescaled distribution of $N_t$ approaches a stationary distribution
\begin{equation}
    W = \lim_{t\to \infty}\frac{N_t}{\mean{N_t}}.
\end{equation}
There are two different ways to derive the distribution of $W$. The standard method for the Bellman-Harris process is to write the moments $a_m\equiv\mean{W^m}$ of $W$ in terms of the Laplace transform of the generation time distribution evaluated at integer multiples of the population growth rate (see \apx{mom_rec} for derivation),
\begin{equation}\label{eq:mom_rec}
    a_1 = 1,\qquad\text{and}\qquad a_m = \frac{\tilde \rho(m)}{1-2\tilde \rho(m)}\sum_{j=1}^{m-1}\binom{m}{j}a_j\, a_{m-j}.
\end{equation}
Substituting for $\tilde \rho(m)$ from \eq{rho_int}, solving the recurrence relationship, then converting moments to cumulants, after a substantial amount of combinatorics calculations, we can find the surprisingly simple expression for the cumulant generating function (CGF)
\begin{equation}
    K_W(p)\equiv \ln\mean{e^{pW}} = \frac{p}{1-\sigma^2p}
\end{equation}
which is the CGF of a compound Poisson-exponential random variable with the density 
\begin{equation}\label{eq:MRD}
    f_W(w) = e^{-\sigma^{-2}}\delta(w) + \frac{e^{-(1+w)/\sigma^2}}{\sigma^2 \sqrt{w}}I_1\left(\frac{2\sqrt{w}}{\sigma^2}\right),
\end{equation}
where $I_1(x)$ is the first-order modified Bessel function of the first kind. The exponentially small term in front of the $\delta$ function is the probability that the first cell (or all of the cells in the population at a later time) gets absorbed at $s=0$ and never divides. 

It is exceptionally rare for the Bellman-Harris process to give a closed-form solution for the asymptotic population distribution. In fact, with the exception of the exponentially distributed generation times (the Yule process), I am not aware of any other generation time distribution that makes BH solvable in closed form. This drastic simplification is not a coincidence. The reason for the simplicity in this model is apparent from \eq{s_tot}, and provides an alternative method that allows us to find the full time-dependent distribution of the number of cells in the population. \Eq{s_tot} shows that the total population mass obeys the same SDE as that of a single cell. While at early times, the number $N_t$ of cells does not track well with the total population mass $M_t$, asymptotically, when the distribution of sizes of the cells settles to its steady state (given in \eq{det_size_dist}), the number of cells is given by $N_t=M_t/\bar s_p$, where $\bar s_p$ is the steady-state population average of cell sizes. The probability density $f_M(s,t)$ of $M_t$ obeys the Fokker-Planck equation
\begin{equation}\label{eq:tot_pop_mass}
    \partial_t  f_M(s,t) = -\partial_s(s\,f_M)+\sigma^2\partial_{ss} (s\,f_M),
\end{equation}
with the initial condition $f_M(s,0)=\delta(s-1)$. These types of PDEs with coefficients linear in $s$ were first studied by Feller~\cite{feller1951two}. Laplace transform reduces this second-order PDE to a first-order PDE, $\partial_t\tilde p(x,t) + (x^2\sigma^2-x)\,\partial_x\tilde p(x,t)=0$, with the initial condition $\tilde p(x,0) = e^{- x}$. This can be solved using the method of characteristics, with the solution $\tilde p(x,t) =\exp[-x\,e^t/(1+(e^t-1)\,x\, \sigma^2)]$. Using the identity $\mathcal L^{-1}\left\{e^{1/s}\right\}=I_1\left(2\sqrt{x}\right)/\sqrt{x} +\delta (x)$ and standard properties of the Laplace transform, we can invert this to find the full time-dependent solution of the probability density 
\begin{equation}
    f_M(s,t) = \exp\left(-\frac{1 + s\,e^{-t}}{\sigma^2(1-e^{-t})}\right)\left(\frac{\sqrt{e^t}}{\sigma^2(e^t-1)}\frac{I_1\left(2\sqrt{\frac{e^t\,s}{\sigma^4(e^t-1)^2}}\right)}{\sqrt{s}}+\delta(s)\right).
\end{equation}
We know that $\mean{M_t} = e^{t}$, so the mean-rescaled variable $U(t)\equiv e^{-t}\,M_t$ would approach a stationary distribution at long time. The density of $U$ is given by
\begin{equation}
    f_{U}(u,t) = f_M(s(u),t) \left|\frac{ds}{du}\right| = \exp\left(-\frac{1 + u}{\sigma^2(1-e^{-t})}\right)\left(\frac{1}{\sigma^2(1-e^{-t})}\frac{I_1\left(2\sqrt{\frac{u}{\sigma^4(1-e^{-t})^2}}\right)}{\sqrt{u}}+\delta(u)\right),
\end{equation}
which reduces to \eq{MRD} at the long time limit. 

We can also recover the prefactor to the exponential growth given in \eq{exp_prefactor} by noting that at the long time limit, 
\begin{equation}
    \mean{N_t} = \frac{\mean{M_t}}{\bar s_p} = \frac{e^t}{\bar s_p},
\end{equation}
and therefore, $c = 1/\bar s_p$. This can be confirmed by directly calculating the average cell size in the population $\bar s_p$ by integrating \eq{det_size_dist} times $s$, recovering precisely the inverse of the expression in \eq{exp_prefactor}. Two useful takeaway identities from this section are the long-time mean and variance of the total number of cells
\begin{equation}
    \mean{N_t} \approx \frac{e^t}{2\ln2-\sigma^2}, \quad\text{and}\quad \frac{\Var{N_t}}{\mean{N_t}^2} = 2\,\sigma^2.
\end{equation}

We have derived these results for a model without division size variability. However, as mentioned in \Sec{pop_growth_rate}, $M_t$ is unaffected by division. Therefore, for a population of cells starting with a single cell of size $s_b = \bar s_b$, regardless of the model of division and cell size regulation, we obtain the same mean-rescaled distribution of population number as in \eq{MRD}. The prefactor (and therefore, the mean population size $\mean{N_t}$), however, is affected by the model of division and cell size control through the population mean $\bar s_p$ of the cell sizes, $\mean{N_t} = e^t/\bar s_p$. Next, we will derive a closed-form expression for the population cell-size distribution in the presence of noise in division size and cell-size regulation, whose mean $\bar s_p$ will give us the full distribution of the number of cells at long time.

\section{Noise in Division Size \& Cell-Size Regulation} \label{sec:div_noise}
So far, we have assumed that all cells are born at size $s_b=1$ and divide at size $s_d=2$. As we will show, the randomness in the division size contributes significantly to the shape of the cell-size distribution. Different cell division strategies have been studied in the context of division size variability and cell size control. A naïve model of cell division, known as the \emph{timer model}, assumes cells would attempt to double their size, such that the division size $s_d$ (conditioned on the birth size $s_b$) is a random variable with the mean $2s_b$. In this case, the log-birth-size of the cell performs a random walk along a lineage and will not converge to a stationary distribution. In practice, cells that are born larger than average tend to grow to less than double their size (on average) before dividing. This process is called cell size regulation. The simplest strategy to do this is to have the distribution of division size $s_d$ be independent of the birth size $s_b$; this is called the \emph{sizer model}. Many biological cells do not do this; instead, they adopt a strategy closer to the simplified model where the added size $\Delta\equiv s_d-s_b$ is independent of the birth size~\cite{amir2014cell, campos2014constant, taheri2015cell}. This is known as the \emph{adder model}. In the context of our model, the cell size control strategy is encoded in how the division size depends on the birth size. We denote the conditional probability density of division size given the birth size by $f^{\rm lin}(s_d|s_b)$. Timer, sizer, and adder models can be represented by choices of $f^{\rm lin}(s_d|s_b)$ such that $s_d-2s_b$, $s_d$, and $s_d-s_b$ are  respectively independent of $s_b$.

Here we do not specify a particular model of cell-size regulation $f^{\rm lin}(s_d|s_b)$. Instead, we assume only that the dependence of $s_d$ on $s_b$ is such that the birth sizes $s_b$ along a lineage converge to a stationary distribution with density $f_b^{\rm lin}(s)$ (there is some mechanism of cell size control, to which we are agnostic). We will show that, given $f_b^{\rm lin}$, the population snapshot cell-size distribution is completely determined and does not depend on the microscopic details of the division noise or size-regulation mechanism that generated $f_b^{\rm lin}$. An analogous decoupling between growth dynamics and size control has been obtained in models with size-independent growth-rate fluctuations~\cite{hein2024asymptotic}.

\subsection{Population Division \& Birth Rates}\label{sec:sym_birth_rate}
In a model with division size variability, the dynamics of cell-size distribution in \eq{det_size_dist_dyn} is modified by source and sink terms coming from birth and division rates at any given size. Therefore, we first need to compute these rates. Let $B_t(s_b)$ be the birth rate density of cells born at size $s_b$ at time $t$. That is, $B_t(s_b)\,dt\,ds_b$ is the number of cells born between size $s_b$ and $s_b+ds_b$ between time $t$ and $t+dt$. Cells born with size $s_b$ come in pairs, from mother cells that divided at a size $2s_b$ that were born at some point $t-\tau$ with the generation time $\tau$, and some birth size we denote by $s_b^-$,
\begin{equation}\label{eq:sym_birth_rate_renew}
\begin{split}    
    B_t(s_b) &= 2\int \int \int B_{t-\tau}(s_b^-)\,\rho(\tau| s_d^-,s_b^-)\,f^{\rm lin}(s_d^-|s_b^-)\,\delta(s_b-s_d^-/2)\,ds_d^-\,d\tau\,ds_b^-
\end{split}
\end{equation}
where $\rho(\tau|s_d,s_b)$ is the conditional density of the generation time at $\tau$ given the birth size at $s_b$ and division size $s_d$. The factor two comes from two daughter cells for each mother cell. At the long time limit, $B_t(s)\propto \beta(s)\,e^{\Lambda t}$, where $\beta(s)$ is the steady-state per-capita birth rate, and $\Lambda=1$ is the population growth rate. Substituting this for $B_t$, we get
\begin{equation}
    \beta(s_b) = 4\int \beta(s_b^-)\, f^{\rm lin}(2s_b|s_b^-)\int e^{-\tau \Lambda}\rho(\tau| s_d=2\,s_b,s_b^-)\,d\tau\, ds_b^-.
\end{equation}
The inner integral is the Laplace transform of the waiting time density for the mother cell starting at the birth size $s_b^-$ to reach $s_d^-=2s_b$ evaluated at $\Lambda$. We have already calculated this Laplace transform in \eq{gen_time_laplace}. Evaluating at $p=\Lambda=1$, we get, 
\begin{equation}\label{eq:gen_EL}
    \mean{\left.e^{-\Lambda T}\right| s_b^-,\,s_d^-} =\frac{s_b^-}{s_d^-}.
\end{equation}
The conditional generation time distribution was the only growth-dependent quantity in \eq{sym_birth_rate_renew}. Given that its Laplace transform evaluated at the population growth rate is given only in terms of $s_b$ and $s_d$, whose lineage distributions are independent of the model of growth, the details of the growth model (namely $\sigma$) drop out of our calculations. After this simplification, we arrive at a simple integral equation for the per-capita birth rate,
\begin{equation}
    \beta(s_b) = \int \frac{2s_b^-}{s_b}\,\beta(s_b^-)\, f^{\rm lin}(2s_b|s_b^-)\, ds_b^-.
\end{equation}
It is straightforward to check that $\beta(s) \propto f_b^{\rm lin}(s)/s$ satisfies this integral equation:
\begin{equation}
    \int \frac{2s_b^-}{s_b}\,\frac{f_b^{\rm lin}(s_b^-)}{s_b^-}\, f^{\rm lin}(2s_b|s_b^-)\, ds_b^- = \frac{2f_d^{\rm lin}(2s_b)}{s_b}= \frac{f_b^{\rm lin}(s_b)}{s_b}.
\end{equation}
This is the second simplification: the integral equation for $\beta(s)$ is written in terms of $f^{\rm lin}(s_d|s_b)$, which encodes the model of cell size regulation, but the solution ends up only depending on $f_b^{\rm lin}(s_b)$, leaving all the subsequent calculations independent of the model of cell size control. 

To fix the normalization, note that the per-capita division rate is the population growth rate $\Lambda$, while the per-capita birth rate is $2\Lambda$. Hence the population per-capita birth-rate density must satisfy $\int_0^\infty \beta(s)\,ds = 2\Lambda=2$, which uniquely determines
\begin{equation}\label{eq:birth_rate}
    \beta(s) = \frac{2\,f_b^{\rm lin}(s)}{s\,\mean{1/s_b}_{\rm lin}}.
\end{equation}
Since cells that divide at size $s$ produce two daughters of size $s/2$, the division rate density $\alpha$ at size $s$ is related to $\beta$ by $2\,\alpha(s)\,ds = \beta(s/2)\,ds/2$, i.e. $\alpha(s) = \tfrac14\,\beta(s/2)$.

In the language of lineage vs tree distributions, this result corresponds to the following: the probability density function of the birth size in the ensemble made of the entire population tree (known as the tree-ensemble) is given in terms of the lineage ensemble by
\begin{equation}
    f_b^{\rm tree}(s) = \frac{1}{\mean{1/s_b}_{\rm lin}}\frac{f_b^{\rm lin}(s)}{s}.
\end{equation}


\subsection{Population cell-size distribution}
\begin{figure}
    \centering
    \includegraphics[width=0.49\linewidth]{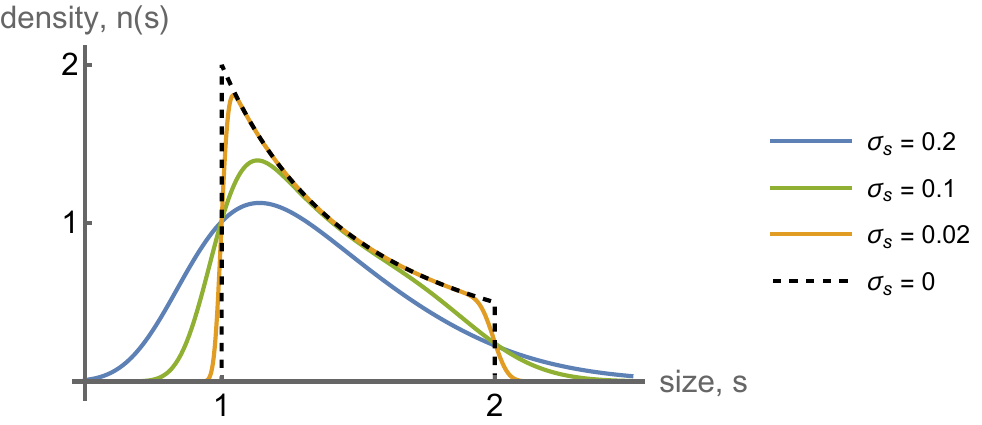}
    \includegraphics[width=0.49\linewidth]{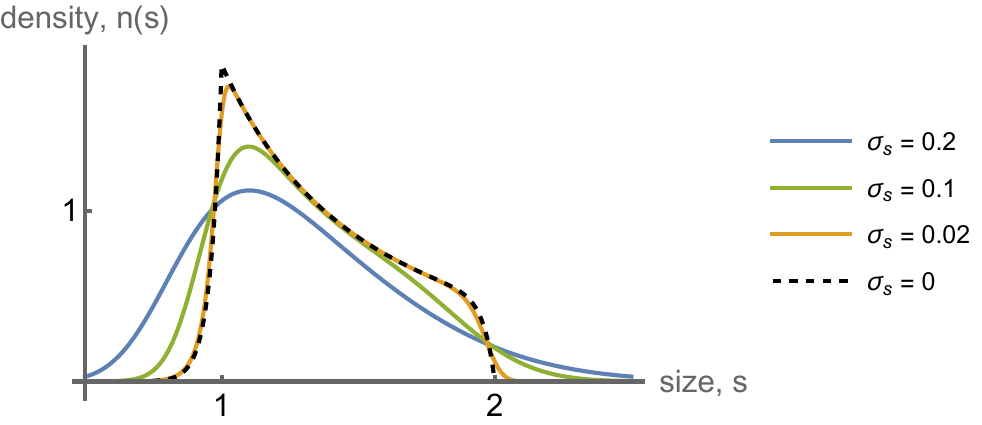}
    \caption{Steady-state population snapshot cell-size distribution for different variability $\sigma_{s_b}$ in lineage birth sizes. The left plot has close to zero growth noise $\sigma=0.01$, while the right plot has growth noise $\sigma=0.2$. The noise in growth effectively changes the cell-size distribution with an exponential moving average with the length scale $\sigma^2$.}
    \label{fig:size_dist}
\end{figure}

The cell-size distribution dynamics with division noise obeys an analogous equation as that of \eq{det_size_dist_dyn}, with the added size-dependent division and birth rates $\alpha(s)$ and $\beta(s)$ as sink and source terms,
\begin{equation}
    \partial_t n(s,t) = -\partial_s(s\,n(s,t)) + \sigma^2\partial_{ss}(s \, n(s,t)) + \beta(s) - \alpha(s) - \Lambda\,n(s,t),
\end{equation}
where the birth rate $\beta(s)$ is given in \eq{birth_rate} and the division rate $\alpha(s)= \tfrac14 \beta(s/2)$. The steady-state solution is given by setting the left-hand side to zero and solving 
\begin{equation}\label{eq:n_SPT}
    \sigma^2 s\, n''(s)+ (2\sigma^2-s)\,n'(s)-2n(s) = \alpha(s) -\beta(s),
\end{equation}
for $n(s)$. There are three ways to solve this equation. The straightforward method is using the Green's function (see \apx{greens}). But to get a more interpretable solution, it is helpful to factor the differential operator on the left-hand side of \eq{n_SPT} as
\begin{equation}
    \underbrace{(-s\,\partial_s-2)}_\text{deterministic transport}\underbrace{(1-\sigma^2\,\partial_s)\, n(s)}_{n_0(s)} = \alpha(s) -\beta(s).
\end{equation}
The operator on the left is the same differential operator we began with (left-hand side of \eq{n_SPT}) with $\sigma$ set to zero. Therefore, $n_0(s)\equiv(1-\sigma^2\,\partial_s)\, n(s)$ is the solution to the deterministic growth model ($\sigma=0$). For deterministic growth, the population cell-size distribution $n_0(s)$ is given by the generalized inverse square law and can be written in terms of the cumulative distribution function of the lineage birth sizes, $F_b(s_b)\equiv\int_0^{s_b} f_b^{\rm lin}(s)\, ds$ as~\cite{koch1966distribution}
\begin{equation}
    n_0(s) = \frac{2}{\mean{1/s_b}_{\rm lin}}\frac{F_b(s)-F_b(s/2)}{s^2}.
\end{equation}
This is the same as the inverse square law, with the discontinuities at $s=1,2$ smoothened out over the range of sizes at which cells are born and divided (see the left panel of \fig{size_dist}). We can solve $n_0(s)=(1-\sigma^2\,\partial_s)\, n(s)$ for $n(s)$ using the integration factor $e^{-s/\sigma^2}$ with the solution
\begin{equation}\label{eq:n(s)_exact}
    n(s) = n_{\rm abs}\,\delta(s)+\frac{2}{\mean{1/s_b}_{\rm lin}}\int_s^\infty \frac{F_b(s')-F_b(s'/2)}{s'^2}\frac{e^{(s-s')/\sigma^2}}{\sigma^2}ds'.
\end{equation}
This is the one-sided exponential moving average of the deterministic growth solution with the length-scale $\sigma^2$. The mass $n_{\rm abs}$ at $s=0$ is of the order of $e^{-1/\sigma^2}$, and is added to ensure normalization. 

The third method for finding the steady-state solution is to use singular perturbation theory~\cite{bender2013advanced} for small $\sigma$ to any order $K$ in $\sigma^2$ as (see \apx{asymp_exp} for derivation)
\begin{equation}
    n(s) = \frac{2}{\mean{1/s_b}_{\rm lin}} \sum_{k=0}^K \sigma^{2k}\frac{d^k}{ds^k}\left[\frac{F_b(s)-F_b(s/2)}{s^2}\right].
\end{equation}
The series can be easily recognized as the series expansion of the operator $(1-\sigma^2\,\partial_s)^{-1}$, and has the exponential kernel with the length-scale $\sigma^2$ given in \eq{n(s)_exact}.

It is useful to evaluate the mean and variance of cell sizes in the population
\begin{equation}
    \mean{s}_{\rm pop} = \frac{2\ln 2}{\mean{1/s_b}_{\rm lin}}-\sigma^2 + O\left(e^{-1/\sigma^2}\right) \quad\text{and}\quad \Var{s}_{\rm pop} = \frac{2}{\mean{1/s_b}_{\rm lin}}-\frac{4(\ln 2)^2}{\mean{1/s_b}_{\rm lin}^2}+\sigma^4 + O\left(e^{-1/\sigma^2}\right),
\end{equation}
where one needs to assume that the lineage density of birth sizes $f_b^{\rm lin}$ decays sufficiently fast near zero. This is a safe assumption, since in practice, biological cells cannot have near-zero birth sizes. There are two interesting observations: (1) the $\sigma$ corrections ($\sigma^2$ shift in the mean and $\sigma^4$ increase in variance) that we saw earlier in \eq{pop_<s>_det} are independent of the division model, and (2) the mean and the variance are affected by the division noise only through the term $\mean{1/s_b}_{\rm lin}$ and are independent of the model of cell size regulation.

\subsubsection{Example: narrow Gamma distributed birth sizes}
The distribution of birth sizes along a lineage can be well-approximated with a narrow Gamma distribution with the mean $\bar s_b=1$ and variance $\sigma_{s_b}^2\ll1$, with the cumulative distribution function
\begin{equation}
    F_b(s) = \frac{\gamma(\sigma_{s_b}^{-2},\sigma_{s_b}^{-2}\, s)}{\Gamma(\sigma_{s_b}^{-2})},
\end{equation}
where $\gamma(a,x)$ is the lower incomplete gamma function.
\Fig{size_dist} shows the population cell-size distribution for this example for a range of $\sigma_{s_b}$ values at near-zero growth noise (Left) and at $\sigma=0.2$ (Right).

The term $\mean{1/s_b}_{\rm lin}^{-1}$ simplifies to $1-\sigma_{s_b}^2$, which gives the mean and the variance
\begin{equation}
    \mean{s}_{\rm pop} \approx 2\ln 2\,(1-\sigma_{s_b}^2)-\sigma^2 \quad\text{and}\quad \Var{s}_{\rm pop} \approx 
    2-4(\ln 2)^2+(8(\ln 2)^2-2)\,\sigma_{s_b}^2-4(\ln 2)^2\,\sigma_{s_b}^4+\sigma^4.
\end{equation}
The division noise further shifts the mean down and increases the variance.


\section{Partitioning Noise \& Asymmetric Division}\label{sec:asym_div}
So far, we have considered symmetric division, where each cell of size $s_d$ divides into two cells of size $s_b = s_d/2$. Now we consider the stochasticity in the division ratio $0<r<1$ such that the two daughter cells have the sizes $s_b=r\,s_d$ and $s_b'=(1-r)s_d$. We assume the division ratio has the probability density $\pi(r)$, which is symmetric around $r=1/2$, and we assume $r$ is independent of $s_d$.

To find the population cell-size distribution, we revisit the decomposition 
\begin{equation}\label{eq:asym_decomp}
    \underbrace{(-s\,\partial_s-2)}_\text{deterministic transport}\underbrace{(1-\sigma^2\,\partial_s)\, n(s)}_{n_0(s)} = \alpha(s) -\beta(s).
\end{equation}
The asymmetric partitioning only affects the population per-capita division and birth rates $\alpha(s)$ and $\beta(s)$. If we can show that both rates are independent of $\sigma$, then $n_0$ satisfying this equation would be truly the $\sigma=0$ solution, and we can reuse this decomposition to write $n(s)$ for $\sigma>0$ in terms of the $\sigma=0$ solution $n_0(s)$.

To find the per-capita birth rate $\beta(s)$, we start with the total birth rate density $B_t(s_b)$ of cells born at size $s_b$ at time $t$. Cells born with size $s_b$ come from mother cells that divided at a size $s_d^-=s_b/r$ that were born at some point $t-\tau$ with the generation time $\tau$, birth size $s_b^-$ and division ratio $r$,
\begin{equation}
\begin{split}
    B_t(s_b) &= 2\int\int\int\int B_{t-\tau}(s_b^-)\,\rho(\tau|s_d^-,s_b^-)\,f^{\rm lin}(s_d^-|s_b^-)\,\delta(s_b-r\,s_d^-)\,\pi(r)\,ds_d^-dr\,d\tau\,ds_b^-
\end{split}
\end{equation}
The only place where $\sigma$ could influence the birth rate equation is through $\rho(\tau|s_d^-,s_b^-)$. Replacing $B_t(s)\propto \beta(s)\,e^{\Lambda t}$, with the population growth rate $\Lambda=1$, we get
\begin{equation}
    \beta(s_b) = 2\int\int\int \beta(s_b^-)\,\underbrace{\cmean{e^{-\Lambda \tau}}{s_d^-,s_b^-}}_{s_b^-/s_d^- \text{ from \eq{gen_EL}}}\,f^{\rm lin}(s_d^-|s_b^-)\,\delta(s_b-r\,s_d^-)\,\pi(r)\,ds_d^-\,dr\,ds_b^-.
\end{equation}
Every element in this integral equation is independent of $\sigma$, so would be its solution $\beta(s)$. The per-capita division rate $\alpha(s)$ is related to $\beta(s)$ by
\begin{equation}
    \beta(s_b) = 2\int \alpha\left(\frac{s_b}{r}\right)\,\frac{\pi(r)}{r}\,dr,
\end{equation}
which makes $\alpha(s)$ also independent of $\sigma$. This allows us to assume $n_0(s)$ satisfying \eq{asym_decomp} is the $\sigma=0$ solution, and we can reuse this decomposition to write $n(s)$ for $\sigma>0$ in terms of the $\sigma=0$ solution $n_0(s)$ as
\begin{equation}
    n(s) = n_{\rm abs}\,\delta(s)+\frac{1}{\sigma^2}\int_s^\infty n_0(s')\,e^{(s-s')/\sigma^2}ds'.
\end{equation}
Thus, even with asymmetric partitioning, the effect of square-root growth noise on the snapshot cell-size distribution is a one-sided exponential smoothing of the deterministic solution $n_0(s)$ with kernel width $\sigma^2$.

To the best of my knowledge, for deterministic growth ($\sigma=0$) with asymmetric division, the analogue of the inverse-square law is not known in closed form in terms of forward (lineage) size statistics. In particular, given a division kernel $f^{\rm lin}(s_d|s_b)$ that leads to a stationary lineage birth-size distribution, no simple expression for $n_0(s)$ in terms of $f_b^{\rm lin}$ alone is available. Recently, however, a closed-form formula has been obtained in terms of backward (retrospective) size statistics~\cite{ocal2025cell}:
\begin{equation}
    n_0(s) = \frac{2}{\mean{1/s_d}_{\rm bck}}\,
             \frac{\mean{r\,F_d^{\rm bck}(s/r)}_\pi - F_d^{\rm bck}(s)}{s^2},
\end{equation}
where $F_d^{\rm bck}(s)$ is the cumulative distribution function of division sizes in the backward lineage ensemble and $\mean{\cdot}_{\rm bck}$ denotes averaging in that ensemble. The backward ensemble is defined by randomly selecting a cell from the steady-state population and tracing its lineage backward in time. In terms of $F_d^{\rm bck}$, the population cell-size distribution for $\sigma>0$ is given by
\begin{equation}\label{eq:asym_n(s)_exact}
    n(s) = n_{\rm abs}\,\delta(s)+\frac{2}{\mean{1/s_d}_{\rm bck}}\int_s^\infty \frac{\mean{r\,F_d^{\rm bck}(s'/r)}_\pi - F_d^{\rm bck}(s')}{s'^2}\,\frac{e^{(s-s')/\sigma^2}}{\sigma^2}ds'.
\end{equation}
For the simple case of symmetric division $\pi(r) = \delta(r-\tfrac12)$, $F_d^{\rm bck}(s)=F_b(s/2)$, and \eq{asym_n(s)_exact} simplifies to \eq{n(s)_exact}.

\section{Discussion \& Future Directions}\label{sec:discussion}

The statistics of single-cell growth are important evolutionary traits that determine the fitness of an organism and contain information about subcellular processes~\cite{holtzman2026disentangling}. Despite the explosion of single-cell measurements over the past decade~\cite{wang2010robust,allard2022microfluidics} and the quantitative characterization of models of division and cell size control~\cite{campos2014constant, amir2014cell, taheri2015cell, si2019mechanistic}, the source and statistics of growth fluctuations remain poorly understood~\cite{thomas2018sources, belliveau2021fundamental}. Growth rate is an emergent quantity, and many subcellular processes contribute to its fluctuations: stochastic gene expression and regulation~\cite{elowitz2002stochastic, paulsson2005models, kiviet2014stochasticity, keren2015noise, scott2010interdependence}, noise in resource allocation~\cite{thomas2018sources, belliveau2021fundamental}, fluctuations in ribosome number and activity~\cite{scott2011growth, lin2018homeostasis}, and variability in cell-wall synthesis~\cite{lee2014cellwall,bissonfilho2017treadmilling}. Asymmetric partitioning of growth-limiting components at division (e.g., ribosomes, enzymes) can induce systematic differences in growth rate between sisters~\cite{thomas2018sources, biswas2024collective, ali2025cyclo}. There is no consensus on the dominant source of noise in cellular growth, and therefore, no agreement on the microscopic description that gives rise to the observed growth fluctuations. From the experimental point of view, extracting high-quality statistics of growth fluctuations remains challenging: ``mother machine'' type experiments can generate growth traces of hundreds of lineages for hundreds of generations each, but the size of a bacterial cell ($\sim 1\,\rm\mu m$) is of the order of the diffraction limit of visible light ($\sim 0.3\,\rm\mu m$), and typical images contain only tens of pixels per cell~\cite{wang2010robust, taheri2015cell, tanouchi2017long}. Small changes in focus and segmentation can alter the inferred size by several percent per frame, and the choice of proxy (length, area, inferred volume, or mass) could change the apparent growth statistics~\cite{Hardo2022SyMBac, tanouchi2017long}. Many mother machine data sets show significant channel-to-channel variability in their lineage statistics, making the characterization of growth statistics even more challenging~\cite{taheri2015cell, elgamel2023multigenerational}. Larger eukaryotic cells are not resolution-limited, but their irregular shapes and complex cell cycles make volume estimation harder~\cite{son2012direct, cadart2018size}. 

All we know with certainty about growth is the following: (1) Most biological cells grow approximately exponentially on average. (2) The exponential growth rate over the cell cycle varies from cell to cell, generation to generation, and within a cell cycle. The within-cell-cycle variability can only be confidently quantified after coarse-graining over a significant fraction of the cell cycle. (3) Large fluctuations in the average exponential growth rate of a cell over its cell cycle decay towards a well-defined lineage mean over the course of a couple of generations~\cite{wang2010robust, Tzur2009, kiviet2014stochasticity, taheri2015cell, biswas2024collective}. Instantaneous single-cell growth rate is estimated by fitting an exponential growth curve to the estimated size from a couple of frames before and after a given time. Typical frame intervals are $\Delta t\sim 1 - 5$ minutes for bacteria and $\sim5 - 20$ minutes for mammalian cells, so “instantaneous’’ growth rates inferred from finite differences already average over many microscopic events and over significant segmentation noise. This coarse understanding of growth fluctuations in single-cell experiments motivates shifting focus towards population-level predictions that can distinguish between different single-cell growth models.

The simplest model to understand the effect of cell-to-cell variability on population dynamics is to assume that each cell has a growth rate drawn from a random distribution that remains constant throughout the cell cycle. In this model, the population growth rate is less than the mean single cell growth rate~\cite{lin2017effects, genthon2025noisy}. However, if growth rates are selected not independently but correlated with the growth rate of the mother cell, there is a crossover to a regime in which growth-rate variability becomes beneficial and increases the population growth rate~\cite{lin2020single}. This is in contrast with the models in which the exponential growth rate $\lambda(t)$ of cells fluctuates continuously (and independently of the cell size) around a well-defined mean, where the growth rate variability is always beneficial and increases the population growth rate~\cite{tuanase2008regulatory, levien2021non, hein2024asymptotic}. Here, by incorporating the square-root dependence of growth fluctuations on cell size, we have obtained an exactly solvable model. 

For experimentally relevant values of $\sigma$, the effect of the particular square-root form of growth noise over the approximately factor-of-two range of cell sizes over the cell cycle is expected to be modest. The main point is therefore not that square-root fluctuations produce a dramatic quantitative change in the observables, but that they lead to a \emph{qualitatively} different population-level decoupling structure: (i) the population growth rate is completely unaffected by the details of growth and division, and is given simply by the mean single-cell growth rate; (ii) the steady-state snapshot cell-size distribution for $\sigma>0$ is obtained by a one-sided exponential convolution with the deterministic $\sigma=0$ solution; and (iii) the mean-rescaled population size $W=N_t/\mean{N_t}$ has the closed-form distribution in \eq{MRD}, which depends only on $\sigma$ for a single founder of size one, and this $W$ is unchanged by division noise, cell size control, and asymmetric partitioning. In models with size-independent single-cell growth rate $\lambda(t)$, growth noise modifies the population growth rate $\Lambda$ but leaves the cell-size distribution $n(s)$ invariant~\cite{tuanase2008regulatory, hein2024asymptotic, levien2025size}. In our square-root growth model, by contrast, growth noise leaves $\Lambda$ unchanged and instead appears in closed form in $n(s)$ and in the fluctuations of $N_t$. A model with size-independent additive growth noise would be structurally different. The total population mass would no longer be governed by the same closed one-dimensional process as in \eq{s_tot}, so the exact solution strategy used here would not carry over. The model with the square-root noise is mathematically the simpler model.

In experimental studies, growth rate variability is often quantified by looking at the cell-to-cell variability $\sigma_\kappa$ in the average growth rate of each cell over its generation time $T$, $\kappa \equiv \ln(s_d/s_b)/T$. To connect the predictions of our model to experiments, it is helpful to express the parameter $\sigma$ in terms of the experimentally measured variability $\sigma_\kappa$ of the cell-cycle-averaged growth rate. In the absence of division noise $\kappa = \ln2/T$ and for small $\sigma$, we can estimate $\sigma_\kappa$ by 
\begin{equation}
    d\kappa = -\frac{\ln(2)}{T^2} dT \implies \left|\frac{d\kappa}{\kappa}\right|= \left|\frac{dT}{T}\right| \implies \frac{\sigma_\kappa}{\kappa}\approx \frac{\sigma_T}{T}\approx \frac{\sigma}{\ln 2},
\end{equation}
where, in the last approximation, we have used \eq{gen_time_mean_var} for the mean and variance of $T$. In the presence of division noise, the first part of this approximation, $\sigma_\kappa/\kappa\approx \sigma_T/T$, does not hold, since the generation time is broadened by the division noise. However, the average growth rate over a period that is slightly broadened by division noise is only affected at order $\sigma_{s_b}^2\,\sigma^2$, which is generally negligible. Hence, it is safe to use 
\begin{equation}
    \frac{\sigma_\kappa}{\kappa}\approx \frac{\sigma}{\ln 2}
\end{equation}
to first-order approximation in $\sigma$ and $\sigma_{s_b}$. Experimentally measured values for growth rate variability in bacteria are typically between $5\%-10\%$~\cite{taheri2015cell}, which translates to $\sigma\sim0.03-0.07$.

A particular quirk of the square-root growth model described in \eq{intro_feller} is that the noise term acts directly on cell size $s$, so over very short time intervals, a cell can shrink. On rare occasions, this shrinkage can drive the cell size to zero, which is an absorbing boundary. We have already shown that the probability of this absorption is astronomically small (of the order of $e^{-1/\sigma^2}$, which is $10^{-90}$ for $\sigma=0.07$); while mathematically important, the absorption has no biological or experimental consequences. However, the bigger concern is whether the short-term shrinking itself is physical, observable, and consistent with experimental data. To address this, we first identify the timescale $\Delta t$ over which the change in the cell size $\Delta s$ is unlikely to be negative. For a cell of size $s=\bar s_b$, we have
\begin{equation}
    \frac{\Delta s}{s} \approx k\,\Delta t+\sigma\sqrt{2k\,\Delta t}\,N <0 \iff N<-\frac{\sqrt{k\,\Delta t}}{\sigma\sqrt2} = -\sqrt{\frac{\Delta t}{\Delta t_c}}, \quad\text{with}\quad \Delta t_c= \frac{2\sigma^2}{k},
\end{equation}
where $N\sim\mathcal N(0,1)$. This means $\Delta s$ is very unlikely to be negative if $\Delta t\gg \Delta t_c = 2\sigma^2/k$. For a bacterial cell with a doubling time of $30$ minutes and $\sigma\sim0.07$, $\Delta t_c\sim 25\,\mathrm{s}$. This is less than the typical 1-5-minute frame rate of mother machine experiments, meaning that any negative growth in cell size in consecutive frames is more likely due to imaging/segmentation errors than actual shrinkage caused by the growth noise. Note that the actual shrinking between frames is at most of the order of $(\Delta s/s)_{\rm max}\sim k\,\Delta t_c=2\sigma^2\sim1\%$, so detecting such negative growth would require sub-percent resolution in single-cell size measurements even with high time resolution, and would in practice be completely masked by imaging and segmentation noise.

Here, we have studied a model of growth fluctuations whose population dynamics is exactly solvable in combination with realistic models of division and size regulation, and which we propose as a tractable benchmark for linking microscopic growth noise to population-level observables. We have shown that this model is (1) a natural one that is (2) not ruled out by the current experimental data, and (3) has population-level properties that are distinct from existing models. Therefore, new experiments are required to probe the nature of growth fluctuations at the single-cell level. In theory, the exact analytical solutions for population dynamics would allow a direct test of this model: one would estimate its parameters from single-cell experiments and then measure population properties such as the growth rate, the population cell-size distribution, or the distribution of the number of cells across many repeated population experiments, comparing these with the analytical predictions. In practice, current technology does not allow for such control over the growth conditions to reproduce the same exact conditions in both single-cell and population experiments~\cite{yang2018analysis}. 

Mutational studies, on the other hand, provide an avenue to test the qualitative differences across different models of growth fluctuations. Conceptually, one can distinguish ``growth-like'' mutants that primarily change the growth process (and thus $k$ and $\sigma$) from “division-like’’ mutants that primarily change division noise and size control (and thus $f^{\rm lin}(s_d|s_b)$ and $f_b^{\rm lin}(s)$) while leaving $k$ and $\sigma$ nearly unchanged. Size-control mutants of the latter type already exist in bacteria and yeast~\cite{scotchman2021identification, yoshida2014directed}. One can use a combination of single-cell measurements (to characterize the effect of mutations) with a population-level measurement to quantify the population observables without the need to quantitatively match all the model parameters across the two experiments. Concretely, a mutation that alters growth statistics (for example, increases the growth fluctuations) would shift the population cell-size distribution under the square-root growth noise, whereas in models where the growth rate fluctuates independently of the cell size the population cell-size distribution would remain unchanged, so comparing these responses across mutants directly discriminates between these two classes of models.

We leave this to future work to study the population dynamics of the extensions of this model such as including correlations and non-Markovian effects~\cite{wei2026subdiffusive}. One useful extension would be to substitute the white noise in \eq{intro_feller} with a colored noise
\begin{equation}   
    \frac{ds}{dt} = k\,s +\sigma \sqrt{s}\,v(t),
    \quad \frac{dv}{dt} = -\theta\,v+ \xi(t).
\end{equation}
This regulates the small-time behavior of the equation and makes the instantaneous exponential growth rate a well-defined quantity, significantly reducing the shrinking problem. At the same time, the auxiliary process $v(t)$ allows us to model temporal correlations in growth rate, representing the memory stored in the concentrations of cellular components in the cell. From the mathematical point of view, it would be interesting to explore other constant-elasticity-of-variance growth models~\cite{pirjol2017phenomenology}
\begin{equation}   
    \frac{ds}{dt} = k\,s +\sigma\, s^\gamma\,\xi(t), \quad 0<\gamma < 1
\end{equation}
in combination with cell size regulation to explore which aspects of our population predictions hold for this broader class of models. Finally, given the importance of segmentation and measurement noise, it would be useful to superimpose a simple error model on the combination of square root noise plus size-regulation dynamics and compute the resulting “observed’’ distributions of cell sizes, growth increments, and population sizes. This would clarify which aspects of the analytical predictions are robust to experimental artifacts, and which are too delicate to be used as quantitative tests.

In summary, the model studied in this paper provides an example in which growth noise is neutral with respect to asymptotic fitness--$\Lambda$ is fixed by $k$--yet still leaves clear signatures in snapshot cell-size distributions and population-size fluctuations. Because these signatures can be written in closed form and disentangle the roles of growth noise and size control, the Feller (square-root) growth plus size-regulation model offers a useful benchmark for interpreting current and future measurements of stochastic growth in size-structured populations.


\appendix
\section{Moment recursion for binary Bellman-Harris}\label{sec:mom_rec}
Consider a population tree starting from a single cell of age zero, each cell either divides into two cells with the probability $p_{\rm div}$ or dies with the probability $1-p_{\rm div}$. Conditioned on division, the generation time density is given by $\rho(\tau)/p_{\rm div}$ (note the expression in the main text for $\rho(\tau)$ is not normalized without the $1/p_{\rm div}$ factor). Let us define the random variable $X$ for each cell that is one if it divides and zero if it dies. Asymptotically, we can write the total number of cells $N_t = W e^{\Lambda t}$ with the population growth rate $\Lambda$ satisfying the Euler-Lotka equation $\mean{e^{-\Lambda T}}=\tfrac12$. 

Now removing the first generation, if the first cell divides, we have two subtrees initiating from the first two daughter cells, each growing for a time $t-T$ would lead to populations of size $N_1 = W_1\, e^{\Lambda (t-T)}$ and $N_2 = W_2\, e^{\Lambda (t-T)}$, which should add up to $N = W e^{\Lambda t}$,
\begin{equation}
    N_1 + N_2  = X\,(W_1+W_2)\, e^{\Lambda (t-T)} = W e^{\Lambda t} \implies W = X\,(W_1 + W_2)\, e^{-\Lambda T}.
\end{equation}
The $m$th moment $a_m \equiv \mean{W^m}$ of $W$ can be recursively computed using this relationship as
\begin{equation}
    a_m = \mean{W^m} = \mean{X\,(W_1 + W_2)^m\, e^{-m\,\Lambda\, T}} = \sum_{j=0}^m \binom{m}{j} \mean{W^j}\,\mean{W^{m-j}}\mean{X\,e^{-m\,\Lambda\, T}} = \tilde \rho(m\Lambda)\sum_{j=0}^m \binom{m}{j} a_j\,a_{m-j}.
\end{equation}
Isolating $j=0$ and $j=m$ terms and solving for $a_m$, we get
\begin{equation}
    a_m = \frac{\tilde \rho(m\Lambda)}{1-2\tilde \rho(m\Lambda)}\sum_{j=1}^{m-1} \binom{m}{j} a_j\,a_{m-j},
\end{equation}
which is the same as \eq{mom_rec} for $\Lambda =1$.

\section{Green's function method for cell-size distribution}\label{sec:greens}
Similarly to the approach in \Sec{det_size_dist}, we define $v(s) = s\, n(s)$ and rewrite \eq{n_SPT} as
\begin{equation}\label{eq:v_gen}
    s\,\sigma^2\,v''(s) - s\,v'(s)-v(s) = s\,(\alpha(s)-\beta(s)),
\end{equation}
with similar boundary conditions as before, $v(0) = 0$, decays at infinity, and the normalization $n_{\rm abs}+\int n(s) ds = 1$, where $n_{\rm abs} = -J(0) = -v(0) +\sigma^2v'(0)$ is the mass absorbed at zero. The homogeneous solutions $v_1$ and $v_2$ are given in \eq{hom_sol}. $v_1$ satisfies the left boundary condition $v(0)=0$ while $v_2$ satisfies the right boundary condition $\lim_{s\to\infty}v_2(s)=0$. A Green's function can be constructed for the equation
\begin{equation}
    v''(s) - \frac{1}{\sigma^2}v'(s)-\frac{1}{s\,\sigma^2}v(s) = \delta(s-s'),
\end{equation}
which has the form
\begin{equation}
    G(s,s') = \frac{v_1(\min\{s,s'\})\,v_2(\max\{s,s'\})}{W(s')},
\end{equation}
where the Wronskian $W(s)\equiv v_1(s)\,v_2'(s)-v_1'(s)\,v_2(s)=-\sigma^2e^{s/\sigma^2}$. The population cell-size distribution is then given by $n(s)=v(s)/s = \int G(s,s') (\alpha(s')-\beta(s'))/\sigma^2\, ds'$. Substituting $\alpha$ and $\beta$ in terms of $f_b^{\rm lin}$, we get
\begin{equation}
    n(s) = n_{\rm abs}\,\delta(s) +\frac{1}{\sigma^4\,s\,\mean{1/s_b}_{\rm lin}}\int_0^\infty v_1\big(\min\{s,s'\}\big)\,v_2\big(\max\{s,s'\}\big)\,\frac{e^{-s'/\sigma^2}}{s'}\,[2f_b^{\rm lin}(s')-f_b^{\rm lin}(s'/2)]\,ds'.
\end{equation}
After some algebraic manipulation and integration by parts, we get
\begin{equation}
    n(s) = n_{\rm abs}\,\delta(s)+\frac{2}{\mean{1/s_b}_{\rm lin}}\int_s^\infty \frac{F_b(s')-F_b(s'/2)}{s'^2}\frac{e^{(s-s')/\sigma^2}}{\sigma^2}ds',
\end{equation}
which is the same as \eq{n(s)_exact}. The mass term $n_{\rm abs}$ at $s=0$ is given by the flux $n_{\rm abs}=-J(0)=\sigma^2\,v'(0)$ which is given by
\begin{equation}
    n_{\rm abs}=-\frac{1}{\sigma^2}\int_0^\infty\left(\sigma^2\,e^{-s/\sigma^2}+s\Ei\left(\frac{-s}{\sigma^2}\right)\right)(\alpha(s)-\beta(s))\,ds.
\end{equation}
With similar manipulations we can rewrite $n_{\rm abs}$ in terms of $F_b(s)$ as
\begin{equation}
    n_{\rm abs} = \frac{2}{\mean{1/s_b}_{\rm lin}}\int_0^\infty \frac{F_b(s)-F_b(s/2)}{s^2}e^{-s/\sigma^2}ds
\end{equation}

\section{Asymptotic expansion of cell-size distribution for small $\sigma$}\label{sec:asymp_exp}

For \eq{n_SPT}, we look for a solution of the form~\cite{bender2013advanced}
\begin{equation}\label{eq:n_ansatz}
    n(s) \sim n_0(s) +\sigma^2 n_1(s) + \sigma^4n_2(s) +\dots.
\end{equation}
Substituting \eq{n_ansatz} in \eq{n_SPT} and collecting terms of similar order in $\sigma^2$, we find the elegant recursive equation
\begin{equation}\label{eq:nk_recursion}
    n_{k+1}(s) = \frac{d}{ds}n_k(s),\quad k\geq0,
\end{equation}
and the leading term $n_0(s)$ is determined by the $\sigma=0$ limit
\begin{equation}
    -sn_0'(s)-2n_0(s) = \alpha(s)-\beta(s) \implies n_0(s) = -\frac{1}{s^2}\int_0^s s'(\alpha(s')-\beta(s'))ds'. 
\end{equation}
Using \eq{birth_rate} for $\beta(s)$ and $\alpha(s) = \tfrac14 \beta(s/2)$, we have
\begin{equation}\label{eq:n0(s)}
    n_0(s) = \frac{2}{\mean{1/s_b}_{\rm lin}} \frac{F_b(s)-F_b(s/2)}{s^2}.
\end{equation}
Combining Eqs.~\eqref{eq:n_ansatz}, \eqref{eq:nk_recursion},  and \eqref{eq:n0(s)}, we find
\begin{equation}
    n(s) = \frac{2}{\mean{1/s_b}_{\rm lin}} \sum_{k=0}^K \sigma^{2k}\frac{d^k}{ds^k}\left[\frac{F_b(s)-F_b(s/2)}{s^2}\right].
\end{equation}

\bibliography{ref2}

\end{document}